\documentstyle[preprint,aps]{revtex}
\tightenlines
\begin{document}
\draft
\title{
Prediction Errors and Local Lyapunov Exponents}
\author{Matthew B. Kennel\cite{MBKauth}}
\address{Institute for Nonlinear Science\\
and\\
Department of Physics\\ 
}
\author{Henry D. I. Abarbanel\cite{HDIAauth}}
\address{
Department of Physics\\ 
and \\ 
Marine Physical Laboratory\\ 
Scripps Institution of Oceanography\\
and\\
Institute for Nonlinear Science}
\author{J. J. (``Sid'') Sidorowich\cite{SIDauth}}
\address{Institute for Nonlinear Science
\bigskip
\bigskip
University of California, San Diego \\ 
Mail Code 0402 \\ 
La Jolla, CA 92093-0402 \\ 
}
\date{\today}
\maketitle
\begin{abstract}

It is frequently asserted that in a chaotic system two initially close
points will separate at an exponential rate governed by the largest
global Lyapunov exponent. Local Lyapunov exponents, however, are more
directly relevant to predictability. The difference between the local
and global Lyapunov exponents, the large variations of local exponents
over an attractor, and the saturation of error growth near the size of
the attractor---all result in non-exponential scalings in errors at
both short and long prediction times, sometimes even obscuring
evidence of exponential growth. Failure to observe exponential error
scaling cannot rule out deterministic chaos as an explanation.  We
demonstrate a simple model that quantitatively predicts observed error
scaling from the local Lyapunov exponents, for both short and
surprisingly long times. We comment on the relevance to atmospheric
predictability as studied in the meteorological literature.

\end{abstract}
\pacs{05.45.+b,92.60.Wc}

\def\de{\mbox{$d_E$}} 
\def\dl{\mbox{$d_L$}} 
\def\l{\lambda}
\def\ol{\mbox{$\bar{\lambda}$}}
\def\u{\mbox{\bf u}}
\def\v{\mbox{\bf v}}
\def\r{\mbox{\bf r}}
\def\x{\mbox{\bf x}}
\def\y{\mbox{\bf y}}
\def\z{\mbox{\bf z}}
\def\F{\mbox{\bf F}}
\def\G{\mbox{\bf G}}
\def\veta{\mbox{$\bf \eta$}}
\def\DF{\mbox{{\bf DF}}} 
\def\DFL{\mbox{$\DF^L$}} 
\def\OSL{\mbox{{\bf O}}} 
\def\doublespace{\parskip 4pt plus 1.5pt
                \baselineskip 20pt plus 1pt minus .5pt
                \lineskip 6pt plus 1pt \lineskiplimit 5pt}

\newcommand{\at}{\mbox{ \ \ \ at\ }}

\newcommand{\lx}{\overline{\lambda}(L)}
\newcommand{\s}{\sigma(L)}

\newcommand{\be}{\begin{equation}}

\newcommand{\bea}{\begin{eqnarray}}
\newcommand{\ee}{\end{equation}}   
\newcommand{\eea}{\end{eqnarray}}   
\newcommand{\dx}{\mbox{$\delta x$}}
\newcommand{\dy}{\mbox{$\delta y$}}

\doublespace
Chaotic behavior in dynamical systems is intermediate between
precisely predictable, regular evolution and completely unpredictable, 
stochastic evolution. The global Lyapunov exponents~\cite{eckru,ose} quantify 
the evolution of perturbations as the system evolves for long times.
A positive global Lyapunov exponent implies that
errors grow to the overall size of the
attractor limiting predictability to finite times.

Given observed data from a chaotic source, one can construct
accurate empirical predictors~\cite{review,farmersid,abk} which allow
predictions for finite times---without any knowledge of the
underlying dynamics. Positive global
Lyapunov exponents causes the errors in these predictions to grow
exponentially rapidly, and it is conventionally assumed that the
prediction error $E(t)$ will grow as
\be
\label{error}
E(t) = E(0) \exp(\l_1 t).
\ee
$\l_1$ is the largest global Lyapunov exponent.
Indeed this error growth is used as a diagnostic of deterministic
chaos in the analysis of observed data~\cite{farmersid,farmersid2}.

This paper is devoted to a closer examination of this common
assumption, and we will show it is often {\em incorrect} for times
where the system is predictable.  For $t \to \infty$ the largest
global Lyapunov exponent $\l_1$ is defined in terms of a long-term
average growth rate, but it does not always reflect the
actual growth of prediction errors in finite time.  Further for $t \to
\infty$ the perturbed orbit, although eventually uncorrelated with the
reference orbit, is constrained to stay on the attractor, and thus the
size of a perturbation will saturate near the overall size of the
attractor.  Reconciling these two aspects of error growth leads to our
improvement to Equation (\ref{error}).

%
%
We briefly review the definition of local Lyapunov exponents.  Our discrete
time dynamical system is $\y(n+1)=\F(\y(n))$.  Small
perturbations $\delta \y(n)$ to this orbit evolve $L$ steps
forward in time according to the
linearized dynamics
\be
\delta \y(n+L) = \DF^L(\y(n)) \cdot \delta \y(n).
\ee
With $\DF_{ab}(\y) =\partial F_a(\y)/\partial y_b$ the Jacobian matrix 
of the dynamics, we denote
$\DF^L(\y(n) = \prod_{i = 0}^{L-1}\,\DF(\y(n + i))$.
The Oseledec matrix~\cite{ose},
\be
\OSL(\y,L)  = \left[ (\DFL(\y))^T \cdot \DFL(\y)
\right]^{1/2L}
\ee
has eigenvalues $e^{\l_1(\y,L)}, e^{\l_2(\y,L)}, \ldots,
e^{\l_d(\y,L)}$, in d-dimensions. We order the exponents as $\l_1
\ge \l_2 \ge \l_3 \ldots \ge \l_d$.
These {\bf local} exponents $\l_a(\y,L)$ address the growth (or decay) over
$L$ time steps of perturbations made around some point $\y$ in the
state space. As $L \to \infty$ $\l_a(\y,L) \to \l_a$, the global
exponents.

Taken over an initially isotropic distribution the initial root-mean-squared
Euclidean distance of the {\em evolved} perturbation vectors to the
origin will grow by $(\frac{1}{d}\sum_{i=1}^d
\Lambda_i^2)^{1/2}$.
$\Lambda_a(\x,L) = \exp[L \l_a(\x,L)]$.  Lorenz first derived this formula 
in an early paper~\cite{lorenz65}.  We employ the
analogous quantity appropriate for a geometric mean instead of the
arithmetic mean. We found this to be $\frac{1}{d}\sum_{i=1}^d
\Lambda_i$ via numerical experiment, though we do not yet know of a
simple, general derivation corresponding to that in
reference~\cite{lorenz65}. We note
that direct numerical computation of the matrix product of many Jacobians
leads to ill-conditioning, and thus for practical computation we employ
the algorithm of reference~\cite{locdat} to stably compute the
local Lyapunov exponents.

The quantity $E(\x,L) =\frac{1}{d}\sum_{i=1}^d
\Lambda_i(\x,L)$, which we denote the ``expansion factor'',
quantifies the multiplicative growth of typical predictor
errors starting at $\x$, looking ahead $L$ steps.  The expansion
factor is best calculated from local exponents as
\bea
\log E(\x,L) &=& L\lambda_1(\x,L) + \log [ 1+
  e^{L(\lambda_2(\x,L)-\lambda_1(\x,L))} + 
  e^{L(\lambda_3(\x,L)-\lambda_1(\x,L))} + \ldots \nonumber \\
  &+&   e^{L(\lambda_d(\x,L)-\lambda_1(\x,L))} ] - \log d  \label{ef}
\eea
For increasing $L$, $\l_1(\x,L)$ quickly dominates
the expansion factor.  This also motivates
Equation~(\ref{error}), but here $\lambda_1(\x,L)$,
is the {\em finite-time} Lyapunov exponent, which does not converge
very quickly to $\l_1$.~\cite{loclap}.

The other ingredient in our model
for the average prediction error is a saturation cutoff.  If we limit
the maximal expansion factor for each initial condition to a constant
$R$, we then compute the geometric mean (over reference
points on the attractor $\x(i)$) of the expansion factors $E(\x,L)$, which is
the arithmetic mean of $\log E(\x,L)$, hard-limited by $\rho=\log R$:
\be
  \log X(L,\rho) = \frac{1}{N} \sum_{i=1}^N \min\left[ \log E(\x(i),L), \rho
\right]
	-\rho. \label{thresheq}
\ee
The saturation cutoff models the fact that finite-sized perturbations
and prediction errors cannot grow indefinitely. We estimate $R$ is the
ratio of the geometric mean of $|\x(i)-\x(j)|$ over uncorrelated pairs
of attractor points to the geometric mean of the initial perturbation
magnitude.  Larger $R$ corresponds to better predictability.  Note
that we are averaging individually thresholded expansion factors---not
thresholding an average expansion factor. Before the saturation sets
in $\log X(L)
\approx L \ol(L)$, with $\ol(L)$ the average finite-time Lyapunov
exponent~\cite{loclap,locdat}.


We now compare (1) the growth of errors actually incurred by making
repeated predictions and (2) the growth rate implied by the
thresholded expansion factors, Equation~(\ref{thresheq}).  This
suggests that prediction errors grow as do the separation of initially
close trajectories.  This means we are considering the errors in
prediction that arise from inherent dynamical instability and not
inaccuracies as a result of specific features of the prediction
scheme.

Our measure of average prediction error is the geometric mean over
initial conditions of an $L$-step iterated predictor, a composition of
$L$ one-step predictors. The error is normalized by the geometric mean
distance between all time-decorrelated pairs of points on the attractor, so 
that the absence of predictability corresponds to zero logarithmic
error. With $\G(\bullet)$ the one-step predictor, the
normalized prediction error $L$ steps ahead is
\be
   \log \chi(L) =  \frac{1}{N} \sum_{i=1}^N \log |\G^L(\x(i)) - \x(i+L)| - 
	\frac{1}{N^2} \sum_{i,j=1}^N 
	\log |\x(i) - \x(j)|. \label{prederr}
\ee
As for $\log X(L,\rho)$ $\lim_{L
\rightarrow \infty} \log \chi(L) = 0$.  The iterated prediction is
{\em not} rescaled to remain close to the reference trajectory.

The main empirical result is that on chaotic attractors $\log \chi(L)
= \log X(L,\rho)$ given the correct threshold $\rho$.  We evaluate the
local Lyapunov exponents, then compute the single parameter family of
curves given by Equation (\ref{thresheq}) over a range of $\rho$.  We
find that once the best value of $\rho$ is found, the resulting $\log
X(L,\rho)$ quantitatively predicts the scaling of errors throughout
the range of time examined.  That equation (\ref{thresheq}) predicts
the scaling of errors in the saturation region ($L \rightarrow
\infty$) is somewhat surprising because Lyapunov exponents quantify
linear expansion rates of {\em infinitesimal} perturbations, but in
the saturation region one envisions typically large deviations.  A
possible explanation is that in the saturation regime average
prediction error is controlled by the tail of the distribution of
individual expansion factors: those initial conditions that happen to
be exceptionally predictable and whose expansion factors thus remain
below the threshold for especially long times.  This is plausible
because local Lyapunov exponents often have wide
distributions~\cite{loclap,locdat}.

Figure~\ref{fig:ikeda} shows iterated prediction
errors,$\log_{10}(\chi(L))$, computed using nonlinear kernel
regression~\cite{abk} versus thresholded expansion
factors (\ref{thresheq}) for data from the Ikeda map of the plane to
itself~\cite{ikeda}. The data set is 20,000 real and imaginary
components from the complex-valued map $z(n+1) = p + B z(n) \exp [ i
\kappa - i \alpha/(1 + |z(n)|^2)]$ with parameters $p = 1.0,\, B =
0.9,\, \kappa = 0.4$, $\alpha = 6.0$.  At short times, there is an
obvious exponential scaling region (a line with a constant slope
$\l_1$), as predicted by Equation~(\ref{error}).  Starting at $L
\approx 10$ the slope decreases and curves off to zero, well modeled
by the thresholded expansion factor.  We did not optimize $\rho$ in
any sophisticated manner, but simply stepped $\rho$ by 0.2 and
selected the best match.  From a time series $\y(i), i=1\ldots N$, we
compute a prediction for an input vector $\x$ as $\G(\x) = \left(
\sum_{i=1}^N
\y(i+1) K(\x-\y(i)) \right) / \left( \sum_{i=1}^N K(\x - \y(i))
\right)$ using a Gaussian kernel $K(\z) = \exp( -
|\z|^2/\sigma^2)$ with $\sigma$ set to twice the geometric mean
nearest-neighbor distance of the $\y(i)$. This is a global prediction
formula but effectively functions as a localized interpolator. The
local Lyapunov exponents were calculated from this same data
set~\cite{locdat} without the equations.

Figure~\ref{fig:lorenzdata} shows the same information for data from a
three dimensional flow of Lorenz~\cite{lorenz84}. The dynamical
equations are
$\dot{x} = - y^2 - z^2 - a(x -F),
\dot{y} =  xy - bxz - y + G,
\dot{z} = bxy + xz - z$, with parameters set $a=0.25, b = 
4.0, F=8.0$, and $G=1.0$, sampled at intervals $\delta t=0.2$.  The
attractor has a dimension of about 2.5. There is no obvious exponential 
scaling regime here.
Given only the curve of predictor errors, one
could not easily identify the Lyapunov exponent, in contrast to the
previous example.  The appropriately thresholded expansion factor,
however, agrees with the observed scaling of prediction errors.  For
contrast, the graph also shows the results of choosing either too
small a threshold (the circles) or too large a threshold (the triangles)
for this prediction scheme and data set.

Now we examine our main result $\log \chi(L)=\log X(L,\rho)$ with a better
controlled experiment.  This time as our ``prediction function'' we
use the actual dynamical equations of the Lorenz system but start the
integration with an initial condition slightly perturbed from the
reference point by a small uniformly distributed random vector
$\veta(i)$: $\G^L(\x(i)) = \F^L(\x(i) + \veta(i))$.  We directly
calculate Lyapunov exponents from the known differential equations by
simultaneously integrating the equations of motion and the tangent
space flow.  Figure~\ref{fig:lorgod} compares $\log \chi(L)$ with
$\log X(L,\rho)$ with varying sizes of the initial
perturbation.  The growth of the perturbations matches the thresholded
expansion factor, with the previously free parameter $\rho$ fixed at
the predicted value $\rho = \langle \log |\x(i)-\x(j)|
\rangle_{i,j=1\ldots N} - \langle \log |\veta(i)| \rangle_{i=1\ldots N}$,
confirming our model.

This scaling with $L$ is generic to most low dimensional
flows: a curve at low $L$ scaling with
an initially high slope, because the average largest
local exponent is larger than the global exponent~\cite{locdat}, an
exponential region (constant slope) corresponding to the global 
Lyapunov exponent, and a long tail curving off to zero as the
finite-size threshold takes effect.  The size and existence of the
region of constant slope depends on the size of the initial error.  If
the error is large enough, there may be no region of exponential
scaling as the curvature due to finite-time exponents merges with the
curvature due to thresholds.  Higher-dimensional data are likely to
have lower initial predictability than very low-dimensional
attractors, and therefore likely to show little observably exponential
error scaling.  Intermittent dynamics will create a wide variation in
finite-time Lyapunov exponents, thus moving forward the time where 
saturation begins to take effect. In analyzing dynamical
systems more complex than simple one or two dimensional models,
requiring manifestly exponential error scaling as confirming
deterministic chaos is unrealistic.

We have identified empirical prediction error with the evolution of
perturbations, but the identity does not always strictly hold.
Figure~\ref{fig:lorllp} shows the prediction errors on the same Lorenz
flow~\cite{lorenz84} data using an accurate local quadratic polynomial
technique~\cite{farmersid,farmersid2}.  This time, the prediction
error is not satisfactorily matched by the thresholded expansion
factor with any $\rho$, the main difference being a substantially
larger slope at small $L$.  We attribute part of this discrepancy to
the assumption, used in deriving the expansion factor, that the
initial perturbations are distributed uniformly in all directions.
When a very accurate model, such as this one, is combined with
extremely clean data, short term forecasting errors will be quite
small, and the predicted trajectory will closely shadow the attractor,
and thus, the unstable manifold.  Therefore the initial rate of
expansion will be larger than for isotropically distributed errors,
being described better by a new ``primary'' expansion factor
$L\lambda_1(\x,L)$ that only considers error expansion due to the
single {\em largest} local Lyapunov exponent.  This quantity increases
faster at short times than the standard expansion factor, but still,
it fails to precisely match the scaling of forecast error.  Another
discrepancy not accounted for by the expansion factor is that
iterating imperfect models injects new error at every timestep, not
only at the beginning.  Usually exponential expansion of old error
dominates new error, except at the shortest times.  This effect will
cause yet another increase to the slope at small times. When we add
some isotropic noise to the initial condition before applying the
iterated local predictor, outstanding agreement with the scaling
predicted by Equation (\ref{thresheq}) is restored.  We note that the
divergence between straight prediction error and expansion factor is
exaggerated by the particular conditions in force here: very good
prediction on a data base of very clean low-dimensional data from a
smooth chaotic flow.  We performed the same test on noisier
experimental chaotic data and saw a smaller difference.

Is adding noise to the initial condition ``cheating''?  To be
completely rigorous, one ought to model the distribution of prediction
errors in the various directions corresponding to the local Lyapunov
exponents.  This is not possible without knowledge of detailed
properties of the specific prediction scheme, and is obviously beyond
the scope of this paper.  The approximation of isotropic perturbation
may often be acceptable, and can be enforced by adding artificial
noise.  In the context of analyzing observed data, a successful match
between prediction errors (even with artificial initial perturbations)
and thresholded expansion factors is a novel cross-check that affirms
the validity of the modeling procedures, even if one cannot exactly
quantify the error scaling for a particular prediction scheme.
Accurate evaluation of Lyapunov exponents from degraded or
high-dimensional data is somewhat difficult in practice, requiring
good estimates of derivatives of the implied evolution function.
Various reconstruction parameters, such as embedding and local
manifold dimensions and time delay~\cite{locdat,review} may yield
substantially different numerical answers for Lyapunov exponents, even
if all are topologically acceptable.  Further requiring a good match
between expansion factors and error scaling provides a criterion for
selecting among the otherwise equivalent choices. In general, we have
found that prediction error scaling varies less with reconstruction
parameters than Lyapunov exponents.

Figure~\ref{fig:expfactoronly} shows $\langle \log E(\x(i),L) \rangle$
($\log X$ without the threshold) for the Lorenz flow.  At the
smallest times, the expansion factor does not in fact possess a
constant slope, and only flattens out to a straight line, with slope
equal to the infinite-time Lyapunov exponent after an initial
transient regime.  This curvature is always in the direction seen on
this graph (higher slope at shorter times) and is a result of the fact
that the average local Lyapunov exponent $\lx$ approaches the global
exponent from above as $L
\rightarrow \infty$\cite{locdat}.  This curvature explains
non-exponential scaling of prediction error for early time intervals.
Also shown is the mean plus and minus one standard deviation of the
distribution of individual expansion factors $E(\x,L)$.  The Figure
demonstrates that the variation in local exponents as a function of
initial condition causes a wide variation in expansion factor that
{\em increases} with increasing $L$.  The interaction of this
wide distribution with the threshold results in the saturation of
prediction error for longer times. The effects of the finite size of
the attractor begin to manifest themselves when an appreciable number
of the individual expansion factors approach the cutoff, which occurs
well before the mean expansion factor does so.  

Notice that the width of the distribution {\em increases} as $L\rightarrow
\infty$; we have observed this feature in all systems we examined.
An important conclusion is that even though the local Lyapunov
exponent converges to a single global exponent which is
{\em independent of initial condition}, i.e. $\lim_{L\rightarrow
\infty} \lambda_1(\x,L) = \lambda_1$, the expansion factors of various
individual initial conditions do not do so: $\lim_{L\rightarrow
\infty} E(\x,L) \ne \bar{E}(L)$.  This is because the standard deviation
of the distribution of local exponents $\langle(\lambda(\x,L) - \lx)^2\rangle$
typically decreases at a rate
substantially slower than $L^{-1}$.  Of course, considering global
saturation, normalized error will eventually converge to unity, but
this is an entirely different mechanism.  The message is that considering
dynamical predictability solely in view of the largest global
Lyapunov exponent may be conceptually misleading as well as quantitatively
inaccurate.

The temporal development of forecast error has been a prime concern of
the weather forecasting community for many years, starting with work
of Lorenz~\cite{lorenz65,lorenz69}.  Common practice has been to
hypothesize {\em ad hoc} relationships for the average error as a function
of time; usually, in the form of a differential equation.  Lorenz
found \cite{lorenz69} that increase of mean deviation between
initially close initial atmospheric states could be fitted by an
empirical law of the form
\be
\dot{E}(t) = \alpha E(t) \left(1-E(t)/E(\infty)\right).
  \label{eq:logistic}
\ee
A positive Lyapunov exponent motivates the first term; the inevitable
saturation at maximum error, the second.  Stroe and Royer~\cite{sr}
compared generalized parameterizations of empirical growth laws that
include Lorenz's with results of large-scale atmospheric simulations
and some experimental observations.  The scaling of mean error at
larger times, in the saturation regime, appeared to be better fit by
an exponential law such as $d\log E(t)/dt = -\beta \log E(t)$, similar
to results seen in our work, but a single empirical rule governing the
initial growth of errors was not clearly indicated, either in this or
previous studies. We explain this with the fact that the initial
growth of error is governed by finite-time, rather than global,
Lyapunov exponents.  This results in an initial regime of
non-exponential growth, second, this error growth rate depends on the
coordinate system used~\cite{loclap}. The differing measures of
phase-space distance employed by atmospheric scientists will result in
different growth rates making the notion of a universal ``doubling
time for small errors'' less useful than commonly believed.  Our
modeling of the error growth, which holds the mean finite-time
Lyapunov exponents responsible at small error, but with their
variation most important at saturation, backs up Stroe and Royer's
conclusions that the infinitesimal growth rate cannot be easily
deduced from the saturation rate via fitting a single empirical
formula like equation (\ref{eq:logistic}).  One further point that we
wish to make is that we have observed that using the arithmetic mean
for the ensemble average of both error and expansion factors rather
than the geometric results in far more ``noisy'' curves, and thus is
it not clear whether expansion factors accurately match prediction
errors with ensemble averaging in the arithmetic sense, though we
suspect so.  The arithmetic mean, commonly employed in meteorological
literature, seems to be dominated by fluctuations in a few samples in
the large error tail of the distribution.  Convergence of the
arithmetical ensemble average appears to require numbers of points
excessively large even for this low-dimensional investigation.
Choosing an arithmetical average also appears to accentuate the
initial superexponential growth.  Still, the qualitative behavior of
error scaling seen in large-scale atmospheric simulations and
experiment~\cite{sr} appears compatible with our model, which we
suggest as a more fundamental explanation for observed error growth.
In the nonlinear dynamics literature, one example in Farmer and
Sidorowich~\cite{farmersid} demonstrated initially
faster-than-exponential error scaling, but remained unexplained in
that work.

With the exception of the work of Lorenz~\cite{lorenz65}, the direct
use of local Lyapunov exponents to quantify error growth in
atmospheric dynamics is rather recent~\cite{farrell,nese,yoden} and
remains open to further development. The Lyapunov exponents have
generally been previously considered only in the context of
infinitesimal errors.  This present work shows that accounting for a
threshold apparently allows finite-time Lyapunov exponents to quantify
error scaling at both small and substantially larger levels of error.
Work remains concerning more realistic models, of course. Results of
other low-dimensional chaotic data sets that we have examined,
including experimental data, agree with the conclusions of this paper.


%
%

\begin{figure}
\caption{ 
$\log_{10} \chi(L)$ (solid line) and $\log_{10} X(L,\rho)$ (circles) for the
Ikeda map. $\rho=2.4$ gives a good fit.}

\label{fig:ikeda}
\end{figure}

\begin{figure}
\caption{$\log_{10} \chi(L)$ (solid line) and $\log_{10} X(L,\rho)$ for data
from the Lorenz flow, with $\rho=1.2$ (triangles), $\rho=1.6$ (circles),
and $\rho=2.0$ (diamonds).}
\label{fig:lorenzdata}
\end{figure}

\begin{figure}
\caption{$\log_{10} \chi(L)$ (solid lines) and $\log_{10} X(L,\rho)$ (symbols)
using known flow equations and varying sizes of initial
perturbations.}
\label{fig:lorgod}
\end{figure}

\begin{figure}
\caption{$\log_{10} \chi(L)$ and $\log_{10} X(L,\rho)$ (symbols)
using local quadratic prediction, with no initial perturbation (solid line),
and a 1\% perturbation (dashed line) before prediction.}
\label{fig:lorllp}
\end{figure}

\begin{figure}
\caption{
Geometric mean of local expansion factors (solid line), plus one
standard deviation of the distribution (dotted line) and minus one
standard deviation (dashed line), for Lorenz flow.}
\label{fig:expfactoronly}
\end{figure}

%
%

\end{document}